\documentclass[12pt]{article}
\usepackage{epsfig}

\parskip=\medskipamount 
plus.3em minus.5em \abovedisplayshortskip=.5em plus.2em minus.4em
\renewcommand{\thesection}{\arabic{section}}

\catcode`@=11

\@addtoreset{equation}{section} \@addtoreset{equation}{subsection}
\def\theequation{\ifnum\value{section}=0 \arabic{equation}\ignorespaces
\else \ifnum\value{section}=-1 A.\arabic{equation}\ignorespaces
\else \ifnum\value{subsection}=0
\thesection.\arabic{equation}\ignorespaces \else
\thesection.\arabic{subsection}.\arabic{equation}\ignorespaces
                             \fi
                        \fi
                   \fi}

{\catcode`\'=\active \def'{{}^\bgroup\prim@s}}

\catcode`@=12

\newcommand{\bq}{\begin{equation}}
\newcommand{\be}{\begin{equation}}
\newcommand{\fq}{\end{equation}}
\newcommand{\ee}{\end{equation}}
\newcommand{\bqr}{\begin{eqnarray}}
\newcommand{\beqs}{\begin{eqnarray}}
\newcommand{\fqr}{\end{eqnarray}}
\newcommand{\eeqs}{\end{eqnarray}}

\def\bop#1{\setbox0=\hbox{$#1M$}\mkern1.5mu
    \vbox{\hrule height0pt depth.04\ht0
    \hbox{\vrule width.04\ht0 height.9\ht0 \kern.9\ht0
    \vrule width.04\ht0}\hrule height.04\ht0}\mkern1.5mu}

\begin{document}
\thispagestyle{empty}

\vskip .6in
\begin{center}

{\bf Computational Enhancement to Programmers}

\vskip .6in

{\bf Gordon Chalmers}
\\[5mm]

{e-mail: gordon$\_$as$\_$number@yahoo.com}

\vskip .5in minus .2in

{\bf Abstract}

\end{center}

Computing according to laymens procedures is changed to contain 
a paradigm of inoptimality in the high level and assembled code.  
The code is changed to maximize the flow of information contained 
in the electrons so that they function more as a group and without 
unwanted coherence effects.  Exponential effects are suspected in 
the improved operation of the programming.  From a laymens point 
of view the maladjustment of substandard code could result in a 
factor of a thousand in such programs as Microsoft Works which can 
be speed intensive.  

\vfill\break 

Modern computer programmers usually take the rule of thumb short 
computer code means fast computer program.  This is an archaic 
philosophy that doesnt take into account the hardware of the computer.  
Typically hardware is a word that means equipment or large piece of 
machinery and in the average laymen or company doesnt realize that 
a small chip is such.  The interwoven flow of data in a single chip 
has not been utilized for various reasons including both the presence 
of danger and the presence of legality.  

For example cohering a billion electrons in such an enviroment can 
cause large wanted and unwanted electromagnetic fluctuations that 
could cause the chip to enhance its program or cause the chip harm 
or even destruction.  Truly an unwanted virus in the form of a 
program could store information in the knucks and krannys of a modern 
chip so that em effects are localized in certain regions.  This 
is to warn the programmer of an altered program that generates 
legality to its use.  

Alternatively the storage of information as it traverses the nanowiring 
of the computer chip can be used to facilitate the speed and memory 
allocation due to the effects of one electron on another or on a local 
group.  Indeed these effects can improve the programs efficiency by 
large factors up to billions or more.  The user code which guides the 
flow of electrons is required to specify the flow of the individual 
electrons.  The actual code that results in the efficiency is not 
necessarily the shortest piece of software that a professional programmer 
is required to present.  

A challenge to a programmer is to write a seemingly unoptimal piece of 
software that is not necessarily the shortest one.  This is complex 
and requires safeguards potentially.  It is also machine dependent 
in that it will function with different quality on one chip than on 
the other.  The computational cost is clearly a function of the degree 
of level of the computer code as it is assembled into machine language.  
The contribution here is to alter the usual computer code into one 
in which the program acts more efficiently without the coherence effect 
which could be shunned and caused by clocking two or more electrons in 
a neighborhood so that they are transmitted more quickly.  Rather the 
code can be designed so that without coherence effects the informaton 
and implementation of the program transmits electrons in a more local 
nature as it is processed through the chip.  The gain in this procedure 
is likely to be a factor of a thousand.  Considering the commonality 
of certain programs such as office programs or machines and programs 
built for fast processing like numerical simulators the improvements 
could result in factors of a thousand or such.  

The implementation of such a program is difficult for a single 
programmer to implement without advanced software tools.  One could 
try compiling the program in billions of possible ways to find an 
optimal program.  Of course this program is not on the market to 
the general public and could result in compiling a program in a 
day instead of seconds.  And this has obvious advantages to the 
user.  Furhermore in a conventional program the memory allocation 
of the various call functions can be rearranged to enhance the 
speed and even accuracy of the program.  

A point in note is the physical flaw in the Intel chip first noticed 
five years ago in that there is a computational bit flaw once in 
$10^9$ of a divide operation.  This can be circumvented perhaps with 
a square root in the frequency of occurence by reassemblying the 
code in a certain form.  

The procedure of guiding the bits through the chip is nonuniversal 
and usually linear as understood by a chip designer who prefer a 
data here to there instead of a how from here to there.  Rather 
a transcendental flow of information within the chip is required 
perhaps even requiring extra bits tagged onto an chirp of data 
to be transmitted so that it affects the flow of information in a 
seemingly inoptimal manner.  This is very nonstandard in typical 
computer programs and appears inoptimal.  However the extra bit 
could cause the memory allocation to occur differently in the 
hardware with a little redundancy.  Even the slightest redundancy 
could exponentiate in the form of $x^q$ with $q$ on the order of  
millions.  Clearly such a procedure is not obvious given the 
linearity of chip structure and the programmers inability to 
analyze a billion versions of seemingly inoptimal code.  There 
are various ways to change the flow of information with nonstandard 
assembly \cite{Chalmers}

However an assembler could be made to fabricate an assembled code 
from typical code that would store and transmit information in the 
best possible manner and also the safest.  Safe here means that the 
program wont overuse certain areas of the chip meaning groups of 
electrons too much.  This is to point out that improving the 
gigahertz processing can be terahertz on your laptop with something 
as simple as using a seemingly counterintuitive programming style 
with the use of various techniques including transcendental operations. 

The use of polytopes can be efficient techique to simulate a chip.  
Large companies could easily optimize their computer programs for 
users with specific computer chips such as AltaVec or Intel or others 
without increasing the coherence effects at all which would result 
in speed ups that could be exponential without any side effects 
to your computer.  Factors of a thousand or more are expected for 
generic programs and should certainly be worth the investment for 
time consuming operating systems such as Windows which is not 
time sensitive but rather for packages such as Maple of Mathematica 
of which there are hundreds of thousands of interfaces.  As a 
anothe point bread and butter packages such as radar operations 
usually optimize the hardware blocks including chips and FPGA 
and ASIC designs with routing of information centralized on point 
to point rather than data specific type that could enhance the 
flow of information through their individual chips.  A modern 
radar could have multiple chips including those of FPGAs or ASICs in 
series and the improvement 
in efficiency which is algorithmic specific will further exponentiate 
which is perhaps more difficult with the field reprogrammable.    

A recommendation is the construction of a random sampling of assemblers 
with things like put an extra bit here or there and further operations 
so that a conventional chip will compile differently and improve both 
the accuracy and efficiency of a generic public program.  It will be 
specific to individual chips designed for the masses and could incorporate 
some higher level operations if the design of the chip is included.  
Perhaps a factor of a million could be achieved in performance speed of 
your desktop application which leads to many man hours wasted removed 
for the user.  The potential payoff truly warrants an investigation 
and construction of a new compiler from big industries.  

It should also be noted that is probabilistically possible to trap 
an electron at a specific site in the chip itself.  This trapping 
is useful for a coherence effect dependent on the program that could 
speed up the electrons by factors of millions.  This is potentially 
dangerous and can be used to deform specific information within the 
memory allocation so that certain types of information are physically 
inaccessible without specific data unsecuring it.  Physically could 
be to garble the information or harm the chip.  Likewise programs could 
be developed against this harm from threat aware viruses.  Programs 
could also be made to operate in extreme environments including 
satellites that encounter specific types of cyclic radiation including 
solar flares or in quantum computers which radiate.

\newpage

\vskip 1 in.

\end{document}